\documentclass[11pt,a4paper]{article}

\usepackage{jcappub}
\usepackage{aas_macros}

\usepackage{graphicx}
\usepackage{bm}

\usepackage{amsmath}
\usepackage{subcaption}
\usepackage{verbatim}

\newcommand{\msolunit}{\mathcal{M}_\odot}

\newcommand{\myeqref}[1]{eq.~\ref{eq:#1}}
\newcommand{\deriv}[2]{\frac{\mathrm{d}#1}{\mathrm{d}#2}}

\newcommand{\rs}{r_\mathrm{s}}
\newcommand{\rhos}{\rho_\mathrm{s}}
\newcommand{\rvir}{r_\mathrm{vir}}
\newcommand{\Mvir}{M_\mathrm{vir}}

\newcommand{\ani}{\beta}
\newcommand{\anilocal}{\ani_\odot}
\newcommand{\sigt}{\sigma_\mathrm{tot}}
\newcommand{\sigr}{\sigma_\mathrm{r}}
\newcommand{\sigrs}{\sigma_{\mathrm{r,s}}}
\newcommand{\sigtlocal}{\sigma_{\mathrm{tot},\odot}}
\newcommand{\sigrlocal}{\sigma_{\mathrm{r},\odot}}
\newcommand{\xlocal}{x_\odot}
\newcommand{\rlocal}{r_\odot}
\newcommand{\rholocal}{\rho_\odot}

\newcommand{\neglogslope}{\gamma}
\newcommand{\ppsdnegslope}{\alpha}

\newcommand{\vr}{v_\mathrm{r}}
\newcommand{\vt}{v_\mathrm{t}}
\newcommand{\vesc}{v_\mathrm{esc}}
\newcommand{\vrp}{v_{\mathrm{r},0}}
\newcommand{\vtp}{v_{\mathrm{t},0}}

\newcommand{\vth}{v_\mathrm{th}}

\title{Derivation of the anisotropy profile, constraints on the local velocity dispersion, and implications for direct detection}
\author{Daniel R. Hunter}
\emailAdd{dhunter@physics.wustl.edu}
\affiliation{Physics Department and The McDonnell Center for the Space Sciences, Washington University, St. Louis, MO 63130, USA}

\abstract{
We study the implications of a pseudo-phase-space density power-law for the anisotropy profile of a Milky Way-like dark matter halo. Requiring that the anisotropy parameter does not take non-physical values within the virial radius places a maximum value on the local radial velocity dispersion. For a plausible range of halo parameters, it is possible to take a local total velocity dispersion of up to about $300\,\mathrm{km/s}$. Making this choice uniquely specifies the anisotropy and dispersion profiles. We introduce a way to model the local velocity distribution that incorporates this anisotropy and study the impact on direct detection.
}

\begin{document}
\maketitle

\section{Introduction}
The shape of the velocity distribution is an important influence on predictions for detection of galactic dark matter (DM)~\cite{Bertone:2004pz}, both indirect~\cite{Robertson:2009bh,Campbell:2010xc,Ferrer:2013cla} and direct~\cite{Gondolo:2002np,Strigari:2009zb,Catena:2011kv,Frandsen:2011gi}. N-body simulations tell us the mass distribution of dark matter in galactic halos, but velocity data is more subject to numerical noise and is thus more difficult to measure. Making assumptions about the phase-space density, one can derive the velocity distribution from a mass density profile~\cite{binney}. This usually involves constraining the anisotropy profile to a certain functional form~\cite{Osipkov:1979,Cuddeford:1991,Gerhard:1991,Baes:2002tw} (alternatively, see~\cite{VanHese:2010qy}). Here we \emph{derive} the anisotropy profile using only information from models of N-body simulations and the Jeans equation (see also~\cite{Zait:2007es}). We impose the physical condition that the anisotropy profile does not rise above one inside the halo, and we discover that this constrains the velocity dispersion profile. In particular we discuss the \textit{maximum value} implied for the local velocity dispersion. We assume halos are spherically symmetric (see~\cite{Sparre:2012zk,Wojtak:2013eia} for studies of velocity anisotropy in aspherical halos) and in equilibrium within their virial radius. It should be kept in mind that while any function that satisfies the collisionless Boltzmann equation also satisfies the Jeans equation, the converse is not necessarily true.

For some years now, it has been apparent that measurements of the pseudo-phase-space density (PPSD) of simulated halos follow a power-law over many decades of radius~\cite{Taylor:2001bq,VanHese:2008ce,Ma:2009ek}. Some early work extrapolated central isotropy everywhere and derived the halo mass distribution from this power-law~\cite{Hansen:2004gs,Dehnen:2005cu,Austin:2005ks}. Work has also been done to explain the dynamical origins of such a power-law.
We will study how assuming a density profile and a PPSD power-law completely specifies the dispersion profile and anisotropy profile of a halo. The sensitivity to the precise slope of the PPSD power-law will be considered. This may be important since evidence for a PPSD power-law so far comes from DM-only simulations. It is becoming viable, however, to include complicated baryonic effects in simulations, which may result in a PPSD slope so far unmeasured or erase the power-law trend completely. Specifically we will use PPSD slopes of $2$, which corresponds to the isothermal profile, $35/18 = 1.9\overline{4}$, the critical value discussed by Dehnen \& McLaughlin~\cite{Dehnen:2005cu}, and $15/8 = 1.875$, the value first found by Taylor \& Navarro~\cite{Taylor:2001bq}. We focus on a Milky Way-sized DM halo, which is specified by the halo parameters: the virial mass $\Mvir$, the scale radius $\rs$, and the concentration $c \equiv \rvir/\rs$, and we use the profile by Navarro, Frenk, and White. 

In Section~\ref{sec:ani_deriv} we outline the derivation of the anisotropy profile from the Jeans equation, the halo profile, and the PPSD profile. Section~\ref{sec:constraints} shows how the anisotropy profile places upper limits on the local velocity dispersion. The anisotropy profile itself is presented in Section~\ref{sec:ani_profile}, accounting for uncertainty in the input parameters. Section~\ref{sec:ani_dist} introduces an anisotropic model for the local velocity distribution, which is used to calculate basic predictions for a generic direct detection experiment. We conclude in Section~\ref{sec:conclusions}, which is followed by appendices containing some details.

\section{Deriving the Anisotropy Profile}\label{sec:ani_deriv}
We use the Navarro, Frenk, and White (NFW) profile~\cite{Navarro:1996gj}. General expressions and some details specific to NFW are deferred to Appendix~\ref{sec:details}. The mass distribution is
\begin{equation}
\rho(x) = \frac{\rhos}{x\left( 1+x \right)^2},
\label{eq:density}
\end{equation}
where $\rhos$ is the scale density and $x \equiv r/\rs$ is the dimensionless radius.
Solving for the contained mass $M(x)$ gives us the scale density $\rhos$ in terms of the virial mass $\Mvir \equiv M(c)$ and concentration $c$.

Following Taylor and Navarro~\cite{Taylor:2001bq} (also see~\cite{Dehnen:2005cu}), we take the PPSD to be a power-law with negative slope~$\ppsdnegslope$:
\begin{equation}
\frac{\rho}{\sigr^3} = \frac{\rho_s}{\sigrs^3} x^{-\ppsdnegslope}.
\label{eq:ppsd}
\end{equation}
The radial velocity dispersion is now known (\myeqref{sigr}), and its value at the scale radius $\sigrs$ may be set by assuming a local radial velocity dispersion $\sigrlocal$.

The anisotropy parameter $\ani$ is defined as
\begin{equation}
\ani \equiv 1 - \frac{\sigma_\mathrm{t}^2}{2\sigr^2},
\end{equation}
where $\sigr$ is the radial velocity dispersion and $\sigma_\mathrm{t}$ is the tangential velocity dispersion\footnote{We define the tangential velocity dispersion such that $\sigma^2_\mathrm{t} = \sigma^2_\theta + \sigma^2_\phi = 2\sigma^2_\theta$.}.
From the integral Jeans equation~\cite{binney}, we can solve for the anisotropy parameter (compare with~\cite{Zait:2007es,Schmidt:2009kz}),
\begin{equation}
\ani(x) = \frac{5}{6}\neglogslope\left(x\right) - \frac{\ppsdnegslope}{3} - \frac{G M(x)}{2x \rs \sigr^2(x)},
\label{eq:ani}
\end{equation}
where we have defined the negative log-log slope of the density $\neglogslope(x) \equiv -\mathrm{d}\log(\rho)/\mathrm{d}\log(x)$.
We know the contained mass $M(x)$, and we know the radial velocity dispersion $\sigr$ from the mass density and PPSD, so we have
\begin{equation}\label{eq:ani_f}
\ani(x) = \frac{5}{6}\neglogslope\left(x\right) - \frac{\ppsdnegslope}{3} - \Sigma^{-2} f(x;\ppsdnegslope),
\end{equation}
where $f(x;\ppsdnegslope)$ is a somewhat complicated function of $x$, with $\ppsdnegslope$ its sole parameter (\textit{i.e.} it does not depend on the halo parameters or $\sigrs$, see \myeqref{nfw_ani} for the full expression).
The quantity $\Sigma$ is a dimensionless measure of the radial velocity dispersion at the scale radius, defined as
\begin{equation}
\Sigma^2 \equiv \frac{\sigrs^2}{4\pi G \rs^2 \rhos /3} = \frac{\sigrs^2}{V_{c,s}^2},
\label{eq:dimensionless_sigmars}
\end{equation}
where $V_{c,s}$ is the circular velocity at the edge of a spherical mass of radius $\rs$ and constant density $\rhos$.

To summarize the necessary ingredients that go into \myeqref{ani_f}, we need the PPSD slope~$\ppsdnegslope$, the halo parameters~$\Mvir$, $\rs$, and $c$ (one of which may be determined by the local halo density~$\rholocal$), the local radial velocity dispersion~$\sigrlocal$, and the local (solar) radius~$\rlocal$.

\section{Constraints from the anisotropy parameter}\label{sec:constraints}
It is shown in Appendix~\ref{sec:ani_limits} that the anisotropy parameter for a NFW profile with PPSD slope $\ppsdnegslope \approx 2$ has asymptotic limits
\begin{equation}
\ani(x) \rightarrow \begin{cases}
  (5-2\ppsdnegslope)/6 & \text{for}\hspace{6pt}x\rightarrow 0\\
  (15-2\ppsdnegslope)/6 & \text{for}\hspace{6pt}x\rightarrow \infty
\end{cases}
\end{equation}
This is acceptable in the small-$x$ limit, where $\ani \rightarrow 1/6$ for, as an example, $\ppsdnegslope = 2$. However, in the large-$x$ limit, with the same value of $\ppsdnegslope$, $\beta \rightarrow 11/6$, which is greater than one, implying an imaginary velocity dispersion. Requiring that $\ani \le 1$ as $x\rightarrow\infty$ would imply $\ppsdnegslope \ge 9/2$, which is a far steeper slope than seen in simulations. This unphysical behavior in $\ani$ may naively suggest that we cannot have a physical model that simultaneously exhibits an NFW density profile and power-law PPSD, but really this requirement for physical-ness is too restrictive. We do not expect the models or assumption of equilibrium (via the Jeans equation) to hold beyond around the virial radius. Requiring that these models are consistent and physical only up to just before they are expected to break down is, however, reasonable and still has consequences elsewhere in a halo. Thus, let us just require that the anisotropy parameter is no greater than one everywhere \textit{within the virial radius}.

Mathematically, we require
\begin{equation}
\forall x\le c : \ani(x) \le 1.
\end{equation}
We can effectively satisfy this for our purposes by requiring that $\ani(c) \le 1$. This gives a maximum value for $\Sigma$ (\myeqref{Sigma2_upper_limit}) that depends only on the concentration $c$ (by way of the virial mass) and PPSD log-slope $\ppsdnegslope$. For reasonable values of $c$ and $\ppsdnegslope$, this upper limit is of order one.
From the definition of $\Sigma$ in \myeqref{dimensionless_sigmars}, this immediately gives an upper bound on $\sigrs$ (\myeqref{sigrs_max}) and thus also on $\sigrlocal$ (\myeqref{sigrlocal_max}) in terms of the halo parameters and $\rlocal$.
\begin{figure}[t]
        \centering
        \includegraphics[width=\textwidth]{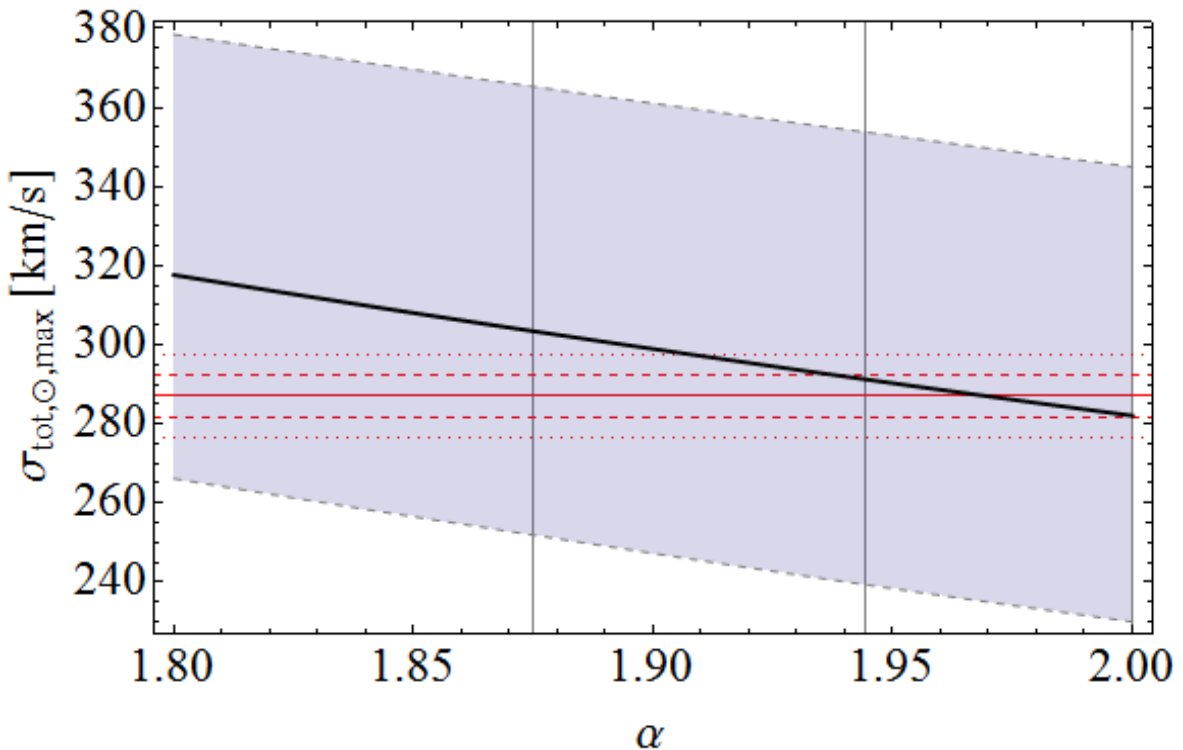}
        \caption{Maximum value of the local total velocity dispersion~$\sigtlocal$. The gray band reflects the uncertainty in the halo parameters: the spread is over the 68\% confidence intervals in Table~\ref{tab:catena_results}. The solid red line marks the mean value for~$\sigtlocal$ found by Catena and Ullio, while the dashed and dotted red lines mark their 68\% and 95\% confidence intervals~\cite{Catena:2011kv}.}
        \label{fig:sigtlocal_max}
\end{figure}
Once $\sigrlocal$ is set, the anisotropy profile $\ani(x)$ is totally specified, including the local anisotropy parameter $\anilocal = \ani(\xlocal)$. The \emph{total} velocity dispersion profile $\sigt$ is then also given, using the relation $\sigt^2 = (3 - 2\ani) \sigr^2$. We find that $\sigtlocal$ depends monotonically on the choice of $\sigrlocal$, so we finally have an upper bound on $\sigtlocal$ (\myeqref{sigtlocal_max}).

\section{Uncertainty in the Maximum Local Dispersion and Anisotropy Profile}\label{sec:ani_profile}
We have derived an upper limit on the local total velocity dispersion $\sigtlocal$, subject to the constraint that the anisotropy profile $\ani(x)$ is no more than one up to the virial radius. This upper limit (\myeqref{sigtlocal_max}) depends on the PPSD slope, the halo parameters, and the solar radius. There is significant uncertainty in these quantities. To get an idea of the uncertainty in the upper limit of $\sigtlocal$, we will use the results of Catena and Ullio~\cite{Catena:2009mf}, summarized in Table~\ref{tab:catena_results}.
\begin{table}[t]
\centering
\begin{tabular}{|c|c|c|c|c|c|}
  \hline
   & lower 95\% & lower 68\% & mean & upper 68\% & upper 95\% \\
  \hline
  \hline
  $\Mvir\;[10^{12}\,\msolunit]$ & $1.23$ & $1.33$ & $1.49$ & $1.64$ & $1.86$ \\
  $c$ & $13.93$ & $16.59$ & $19.70$ & $22.90$ & $24.6$ \\
  $\rholocal\;[\mathrm{GeV/cm^3}]$ & $0.338$ & $0.365$ & $0.389$ & $0.414$ & $0.435$ \\
  $\rlocal\;[\mathrm{kpc}]$ & $7.67$ & $8.00$ & $8.28$ & $8.55$ & $8.81$ \\
  $\sigtlocal\;[\mathrm{km/s}]$ & $276.7$ & $281.7$ & $287.0$ & $292.2$ & $297.2$ \\
  $\vesc\;[\mathrm{km/s}]$ & $528.5$ & $539.7$ & $550.7$ & $561.7$ & $573.3$ \\
  \hline
\end{tabular}
\caption{Assumed ranges for the halo parameters, solar radius, local total velocity dispersion, and local escape speed. Taken from Table~3 of~\cite{Catena:2009mf} and Table~1 of~\cite{Catena:2011kv}.}
\label{tab:catena_results}
\end{table}
With these ranges of parameters, we plot the upper limit of $\sigtlocal$ versus the PPSD slope $\ppsdnegslope$ in Figure~\ref{fig:sigtlocal_max}. The dark, solid line uses the mean values in Table~\ref{tab:catena_results}, while the upper and lower dashed lines take the extreme values of $\sigtlocal$ allowed by the 68\% confidence intervals in Table~\ref{tab:catena_results}. In other words, the band in Figure~\ref{fig:sigtlocal_max} includes all combinations of parameters within the 68\% confidence intervals. This is one of our main results. Also shown is the mean value and 68\% and 95\% confidence intervals for $\sigrlocal$ in~\cite{Catena:2011kv}.

Actually choosing a value for $\sigrlocal$ (or $\sigtlocal$) determines the anisotropy profile, but this quantity is also uncertain. We use the results for $\sigtlocal$ from~\cite{Catena:2011kv} and then take $\sigrlocal^2 = \sigtlocal^2/3$, which is used to find the anisotropy profile in \myeqref{ani}. Note that the factor $1/3$ corresponds to the isotropic case. As we will see, we find only radial bias at the solar radius. Given the same value of $\sigtlocal$, radial bias implies a larger value of $\sigrlocal$, which in turn gives a greater local radial bias\footnote{Successive adapting of the relation between $\sigrlocal$ and $\sigtlocal$ would, of course, converge to the correct ``trial~value'' for~$\anilocal$.}. So as far as predicting departure from isotropy, this is a conservative approximation.

\begin{figure}[t]
        \centering
        \includegraphics[width=\textwidth]{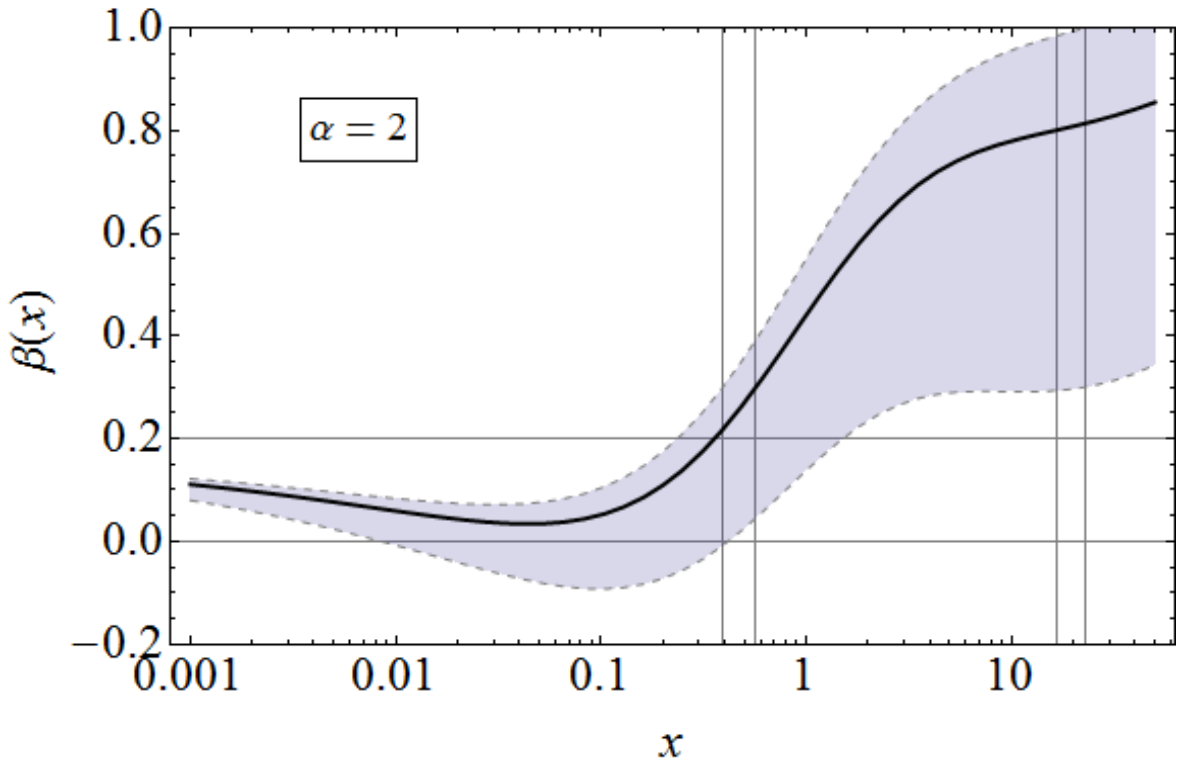}
        \caption{The anisotropy profile for $\ppsdnegslope = 2$, corresponding to the isothermal case, as a function of $x\equiv r/\rs$. The gray band reflects the uncertainty in the halo parameters: the spread is over the 68\% confidence intervals in Table~\ref{tab:catena_results}. The two pairs of vertical lines represent the 68\% confidence interval for the local radius $\rlocal$ and halo scale radius $\rs$.}
        \label{fig:ani_profile2}
\end{figure}
\begin{figure}[t]
        \centering
        \includegraphics[width=\textwidth]{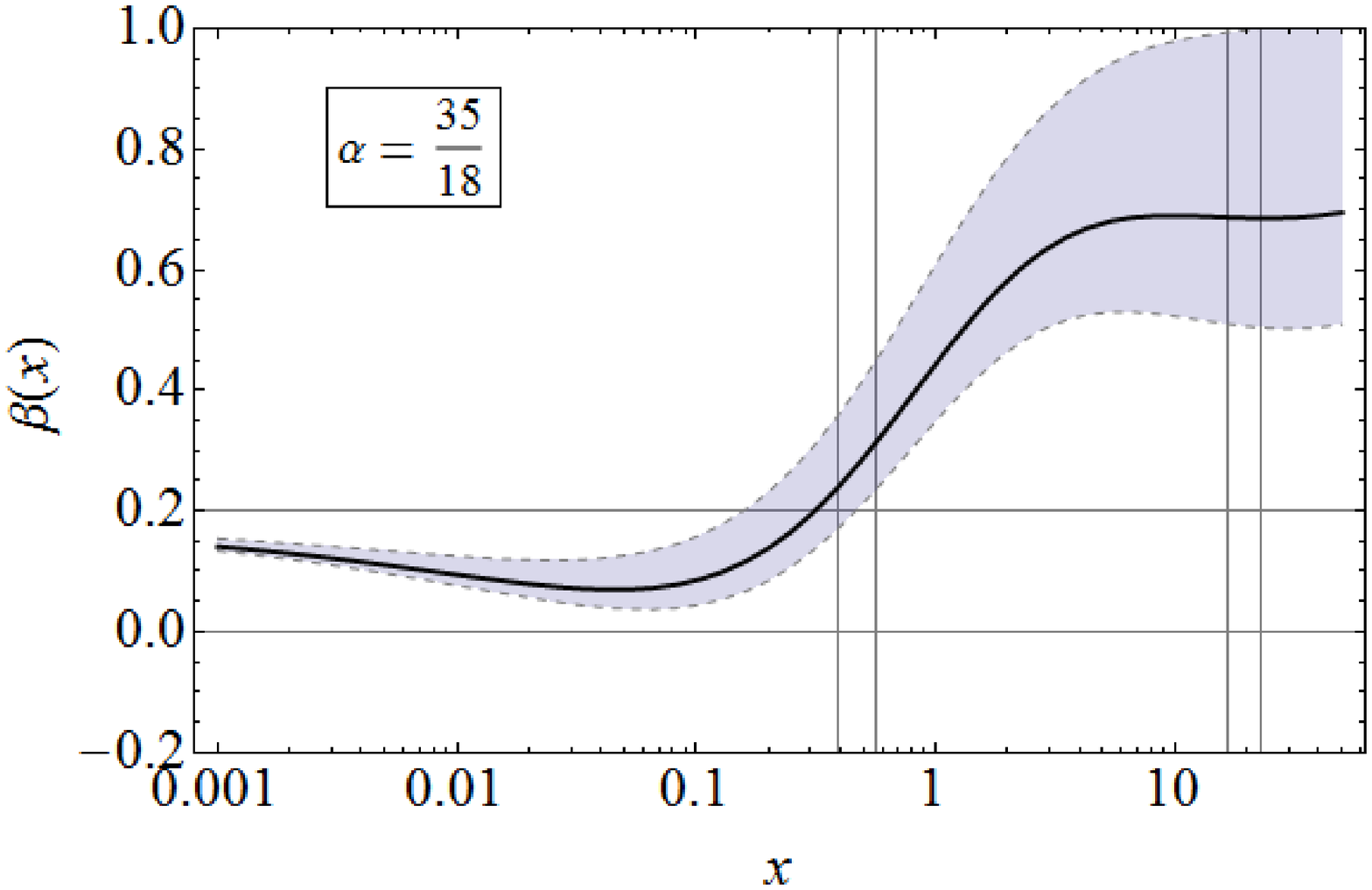}
        \caption{The anisotropy profile for $\ppsdnegslope = 35/18$, the critical value discussed in~\cite{Dehnen:2005cu}, as a function of $x\equiv r/\rs$. The gray band reflects the uncertainty in the halo parameters: the spread is over the 68\% confidence intervals in Table~\ref{tab:catena_results}. The two pairs of vertical lines represent the 68\% confidence interval for the local radius $\rlocal$ and halo scale radius $\rs$.}
        \label{fig:ani_profile35_18}
\end{figure}
\begin{figure}[t]
        \centering
        \includegraphics[width=\textwidth]{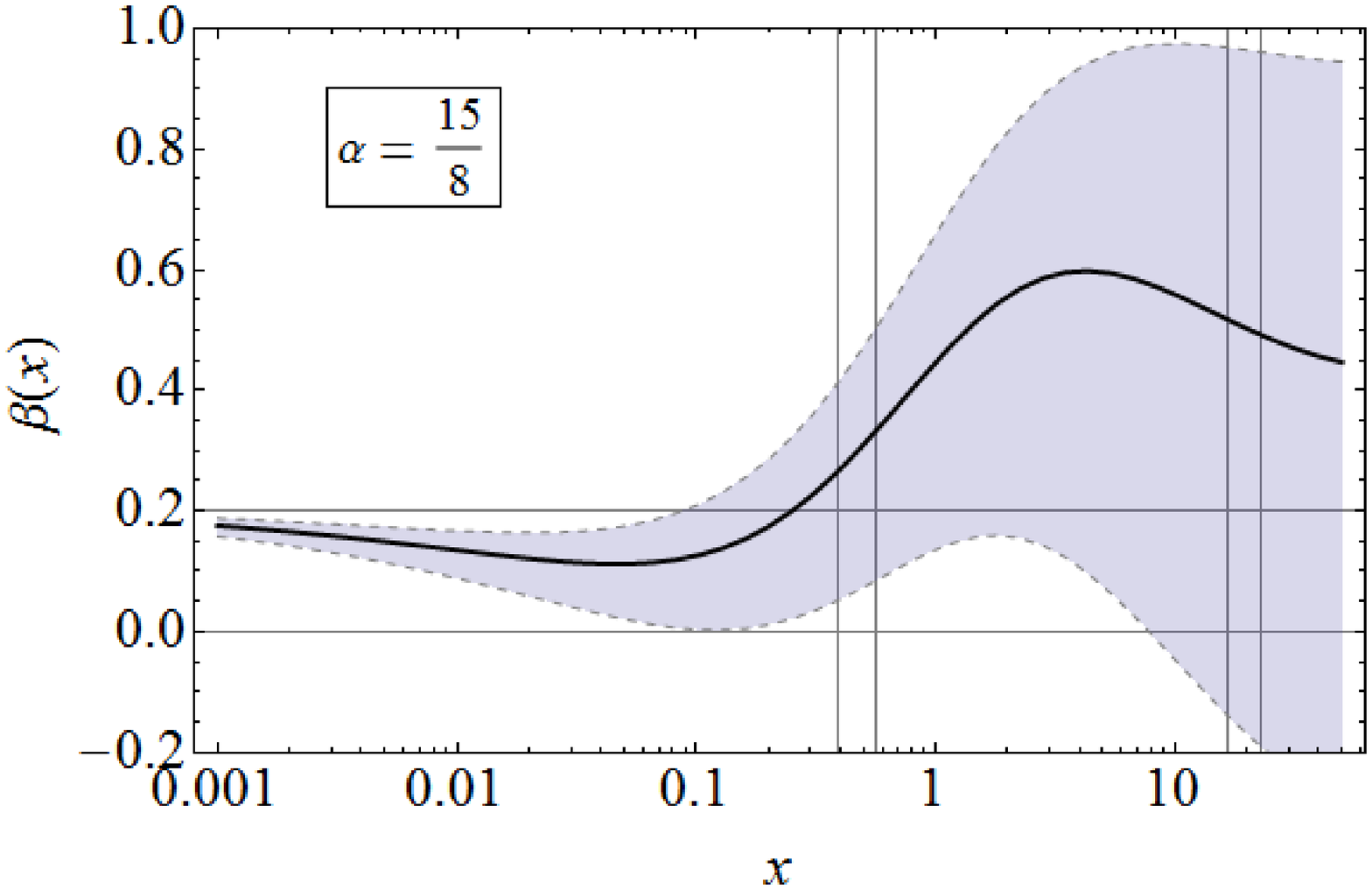}
        \caption{The anisotropy profile for $\ppsdnegslope = 15/8$, the value found in~\cite{Taylor:2001bq}, as a function of $x\equiv r/\rs$. The gray band reflects the uncertainty in the halo parameters: the spread is over the 68\% confidence intervals in Table~\ref{tab:catena_results}. The two pairs of vertical lines represent the 68\% confidence interval for the local radius $\rlocal$ and halo scale radius $\rs$.}
        \label{fig:ani_profile15_8}
\end{figure}
We plot the anisotropy profile for fiducial values $\ppsdnegslope = 2,35/18, 15/8$ in Figures~\ref{fig:ani_profile2},~\ref{fig:ani_profile35_18}, and~\ref{fig:ani_profile15_8}. The solid curve takes the mean values in Table~\ref{tab:catena_results} while the dashed curves are the extreme cases, with all parameters within the 68\% confidence interval in Table~\ref{tab:catena_results}. The vertical lines mark the 68\% lower and upper limits of $\xlocal = \rlocal/\rs$ and $c$. For example, if we assume a PPSD slope of $35/18$ (Figure~\ref{fig:ani_profile35_18}), we might expect a local anisotropy parameter of at least approximately $0.2$ and no more than about $0.4$.

Generally, the profile is slightly radially biased near the center, reaches a minimum at around a tenth the scale radius, and rises to a (local) maximum of around $0.4$ to $0.6$ before the virial radius.
We see in all cases that for $x\rightarrow 0$ the anisotropy parameter rises slowly to the value in \myeqref{Aasym}, which is independent of the halo parameters. See~\cite{Hansen:2004qs,An:2009nc} for discussion of central anisotropy. Here we do not presume that either assumed model, of the mass distribution or PPSD, necessarily stays valid at very small or very large radii. See~\cite{Ludlow:2011cs} for an investigation of the break-down of the PPSD power law.

\section{Anisotropic velocity distributions and predictions}\label{sec:ani_dist}
Recently, close attention has been paid to the form of the velocity distribution used to calculate predictions for indirect and direct DM detection. In some cases the functional form can make a significant difference. Especially, the assumed velocity distribution influences the interpretation of results from direct detection experiments~\cite{Ullio:2000bf,Vogelsberger:2008qb,Catena:2011kv}. Here on we focus on the local distribution and suppress the subscript $\odot$. We introduce a new, anisotropic generalization of the model proposed by Mao, et al.~\cite{Mao:2012hf}:
\begin{equation}
f(\mathbf{v}) \propto \mathrm{exp}\left\{-\sqrt{\frac{\vr^2}{\vrp^2} + \frac{\vt^2}{\vtp^2}} \right\} \left(\vesc^2 - v^2 \right)^p,
\label{eq:animao}
\end{equation}
where $\vr = v\cos\left(\eta\right)$ and $\vt = v\sin\left(\eta\right)$ are the radial and tangential velocity components and $\eta$ is the angle from the radial direction. The parameters $\vrp$ and $\vtp$ are not dispersions but just velocity scales. The exponent $p$ characterizes the high-velocity tail. The function is normalized so that $\int \mathrm{d}\mathbf{v} f\left(\mathbf{v}\right) = 1$.
We choose this distribution because of its recent success in modeling the Eris simulation (see Fig. 3 in~\cite{Kuhlen:2013tra}). For consistency with that study we take $p=1.5$, which was used to model the ErisDark results\footnote{We do not take the Eris parameter $p=2.7$ for two reasons: we have not considered baryonic effects on the PPSD profile, and because such a steep cut-off makes it difficult to achieve anisotropy greater than $\ani\approx 1.0$ with the model in \myeqref{animao}.}. The escape speed $\vesc$ is given by the combined gravitational potential of both the DM halo and any other matter, and we use the mean value in Table~\ref{tab:catena_results}. The total dispersion $\sigt$ and the anisotropy parameter $\ani$ are then determined by the parameters $\vrp$ and $\vtp$. We require that the total dispersion equals the mean value in Table~\ref{tab:catena_results} and solve for $\vrp$ and $\vtp$ such that the desired anisotropy parameter is generated. Of course, the original, isotropic distribution is recovered when $\vrp=\vtp$. See Appendix~\ref{sec:vd_details} for details on the selection of values for $\vrp$ and $\vtp$.

We have checked that the uncertainties in the values of $\sigt$ and $\vesc$ have a small impact on the following calculations. More importantly, the uncertainties affect both the isotropic and anisotropic cases equally once $\ani$ has been chosen. So for the purposes of investigating the importance of modeling deviation from isotropy, we show only results using the mean values in Table~\ref{tab:catena_results}.

\begin{figure}[t]
        \centering
        \includegraphics[width=\textwidth]{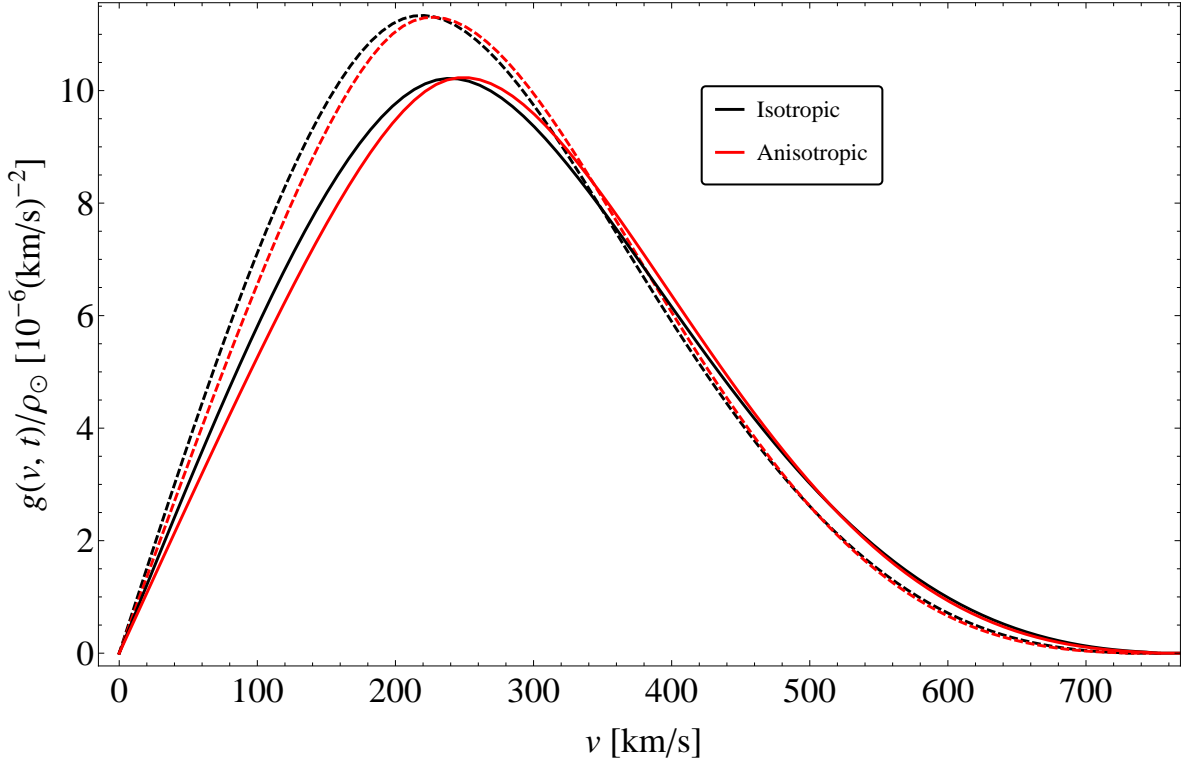}
        \caption{The function $g\left(v,t\right)$ defined in \myeqref{g}. Solid lines are calculated in June; dashed lines are calculated in December.}
        \label{fig:g}
\end{figure}

We use the function in \myeqref{animao} to model the local velocity distribution with the intention of understanding the impact that anisotropy can have on direct detection. For the purposes of this work, we assume a conservative value of $0.2$ for the anisotropy parameter $\ani$. It is straight-forward to calculate the function
\begin{equation}
g(v,t) = \rho_\odot \int_0^\pi \mathrm{d}\eta\sin\left(\eta\right) \int_0^{2\pi}\mathrm{d}\psi\,v f\left(\mathbf{v}_\mathrm{halo}\right),
\label{eq:g}
\end{equation}
where $\mathbf{v}_\mathrm{halo}$ is the velocity vector boosted from the detector frame to the halo frame~\cite{Catena:2011kv}. The boost depends on the time of the year $t$. In Figure~\ref{fig:g} we plot this function for June and December; for the isotropic case and the anisotropic case.
Using the function $g\left(v,t\right)$, the differential detection rate is found by specifying a velocity threshold $\vth$ for DM particles in the detector frame:
\begin{equation}
\deriv{R}{Q} \propto G\left(\vth, t\right) \equiv \int_{v \ge \vth} \mathrm{d}v\,g\left(v,t\right).
\label{eq:rate}
\end{equation}
\begin{figure}[t]
        \centering
        \begin{subfigure}[b]{0.48\textwidth}
                \centering
                \includegraphics[width=\textwidth]{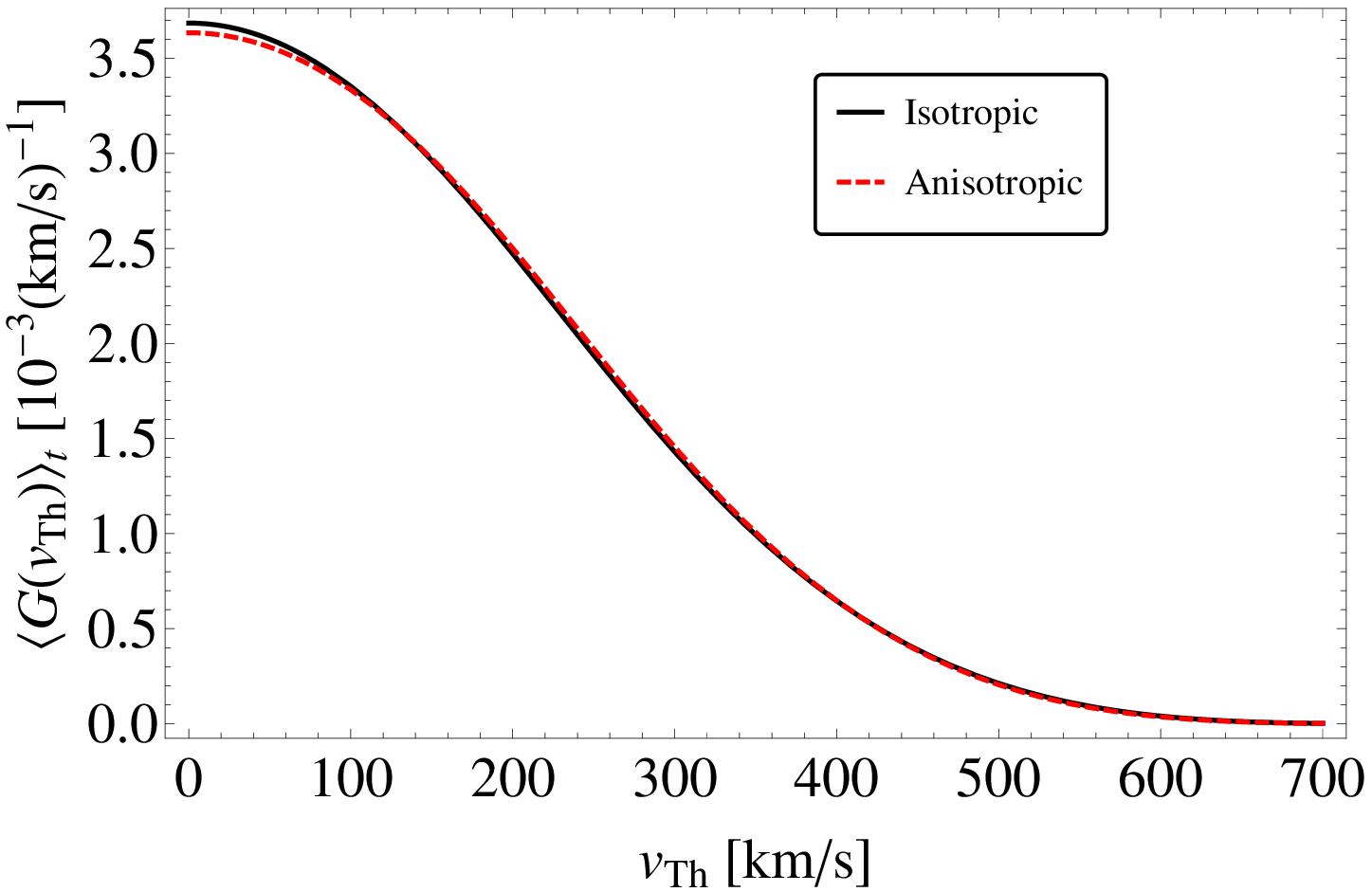}
        \end{subfigure}
        \begin{subfigure}[b]{0.48\textwidth}
                \centering
                \includegraphics[width=\textwidth]{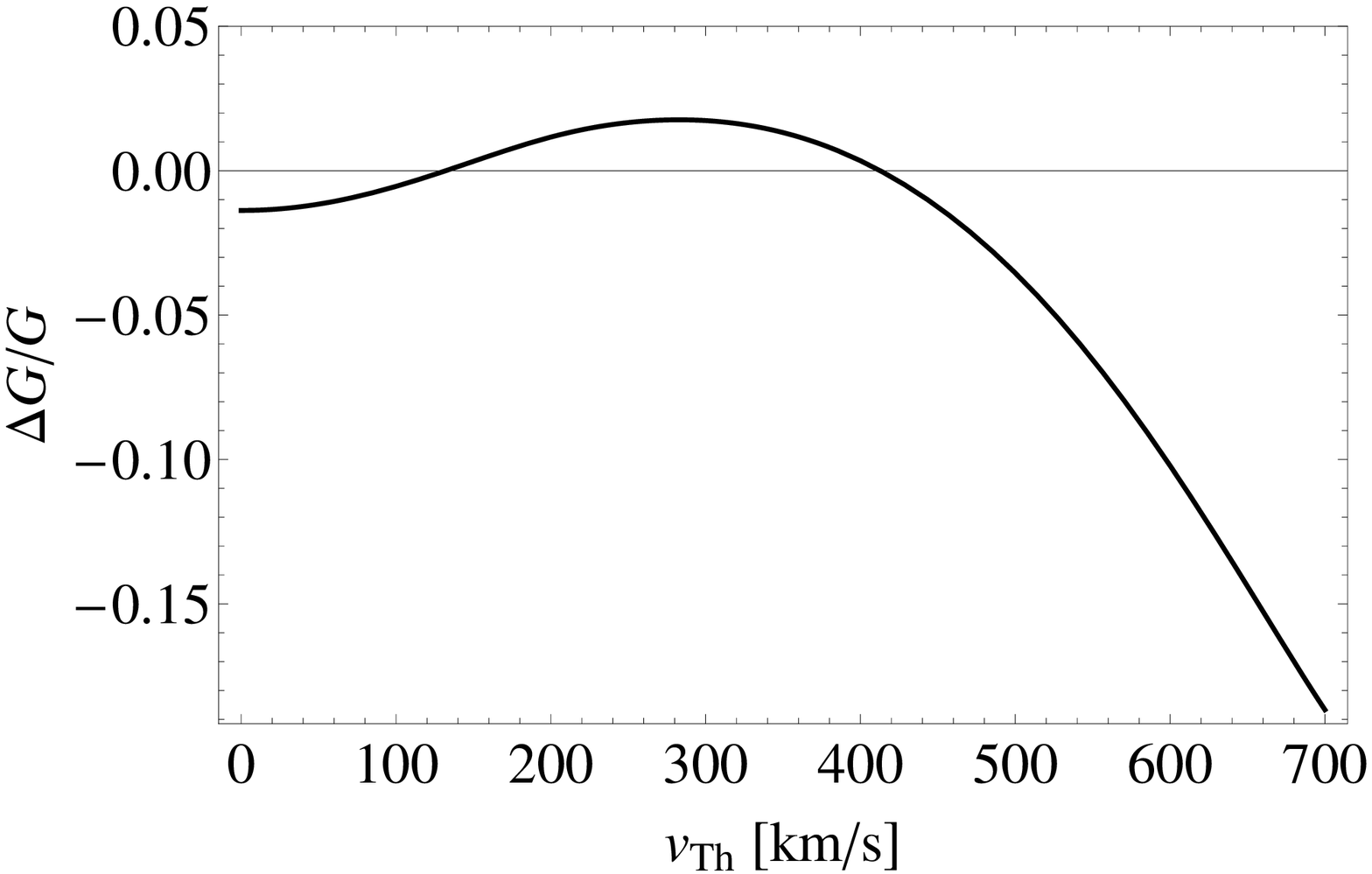}
        \end{subfigure}
        \caption{The time-averaged function $\langle G\left(\vth,t\right)\rangle_t$ as a function of the velocity threshold (see~\myeqref{rate}) and the fractional difference between the isotropic to anisotropic cases.}
        \label{fig:rate}
\end{figure}
\begin{figure}[t]
        \centering
        \begin{subfigure}[b]{0.48\textwidth}
                \centering
                \includegraphics[width=\textwidth]{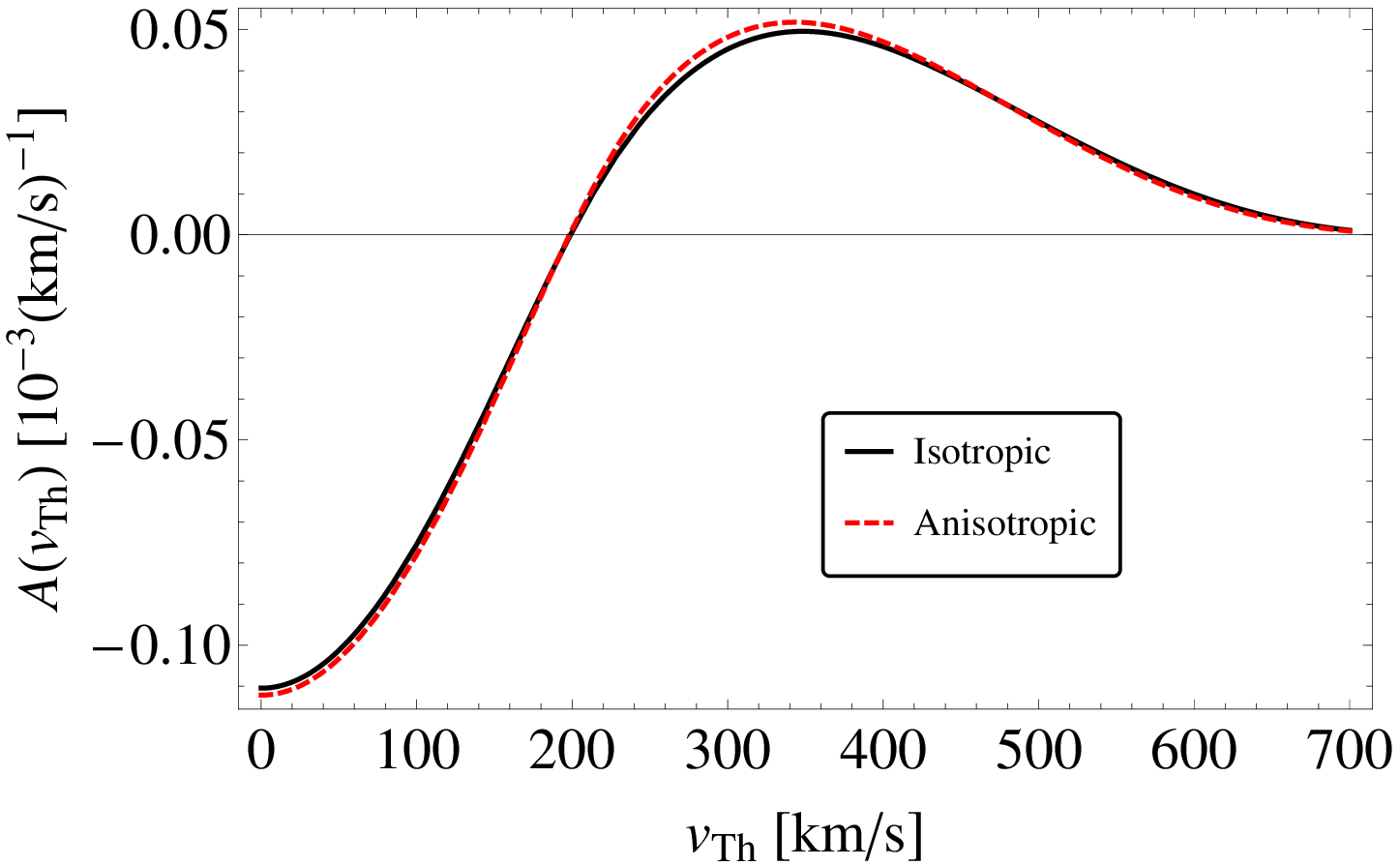}
        \end{subfigure}
        \begin{subfigure}[b]{0.48\textwidth}
                \centering
                \includegraphics[width=\textwidth]{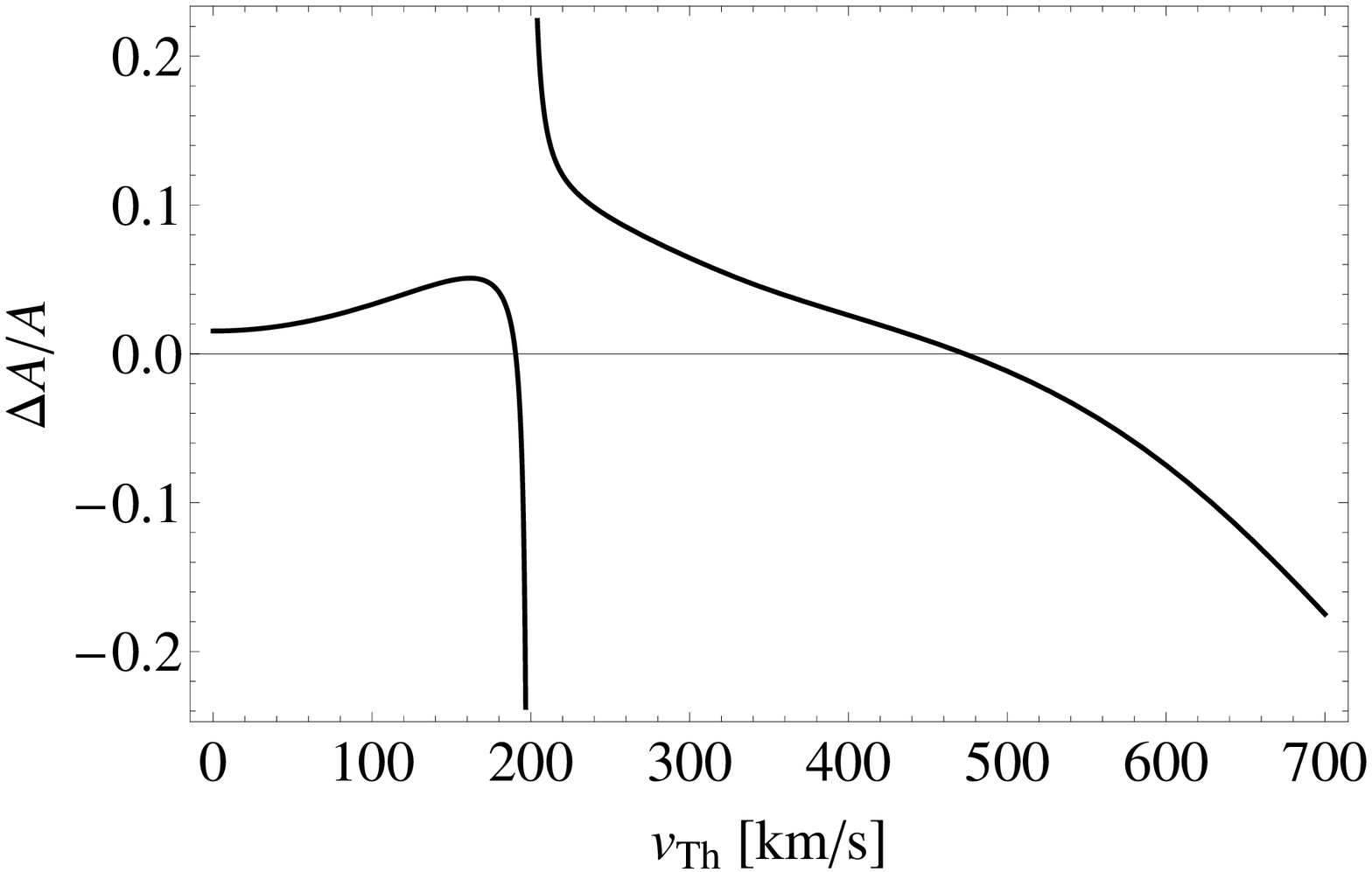}
        \end{subfigure}
        \caption{Signal modulation amplitude as a function of the velocity threshold and the fractional difference between the isotropic to anisotropic cases.}
        \label{fig:mod}
\end{figure}
The velocity threshold is determined by the specifics of any particular experiment and the DM particle mass, and we leave it free. Figure~\ref{fig:rate} plots the function $G$, averaged between June and December, for the isotropic and anisotropic cases, with the fractional difference
\begin{equation}
\Delta G = \left( G_\mathrm{Ani} - G_\mathrm{Iso}\right)/G_\mathrm{Iso}.
\label{eq:G_frac_diff}
\end{equation}

We also consider the modulation amplitude of the signal, defined here as half the difference between the rate in June and the rate in December:
\begin{equation}
A\left(\vth\right) = \left| G\left(\vth,t_\mathrm{June}\right) - G\left(\vth,t_\mathrm{Dec}\right)\right|/2.
\end{equation}
This is plotted in Figure~\ref{fig:mod} for the isotropic and anisotropic cases, with the fractional difference, analogous to \myeqref{G_frac_diff}.

\section{Conclusions}\label{sec:conclusions}
Combining models of the mass distribution and pseudo-phase-space density, the Jeans equation gives us a particular anisotropy profile. We have plotted this profile for a few representative values of the PPSD slope and for a spread of parameters that describe the galactic halo. These profile shapes are consistent with those shown in Figure~1 of~\cite{Zait:2007es}, although those results exhibit less anisotropy overall. The anisotropy profiles found in~\cite{Bozorgnia:2013pua} are also similar but were derived from models of the phase-space distribution. The difference in methods strengthens both their results and these.

We have used an anisotropic modification to the model proposed by Mao et al.~\cite{Mao:2012hf}, which was also used to model the Eris simulation. We find that assuming a local anisotropy of approximately $0.2$ is reasonable and conservative. In the Eris simulation, a comparable amount of radially biased anisotropy was found at the location corresponding to the solar radius (this is roughly seen by measuring the half-maximum width of the radial and azimuthal distributions in Figure~2 of~\cite{Kuhlen:2013tra}). On the other hand, the results of~\cite{Catena:2009mf} favor a local \textit{tangential} bias, though the small local radial bias found in this work and others already mentioned is approximately within their 95\% confidence interval.

Different direct detection collaborations have found contradictory results (\emph{e.g.} see~\cite{Gondolo:2012rs}). Part of the general goal in studying the local velocity distribution is to alleviate these discrepancies. Since different experiments can have different threshold velocities, Figure~\ref{fig:rate} suggests that the difference between observed signals can vary by several percent due to the effect of local anisotropy. This may seem small, but it is comparable to the uncertainty introduced by considering different density profiles~\cite{Catena:2011kv}. The modification to the modulation amplitude can be even more significant and is sensitive to the value of the velocity threshold, but the signal itself is smallest where the modification is greatest.

In principle, a detector that can give information about the direction of a detected WIMP's velocity would allow us to \emph{measure} the local anisotropy. This is difficult, as it would require an individual WIMP to interact multiple times inside the detector or require a low detector density so the recoiled particle can be tracked. Once a discovery is confirmed, however, it may be viable to consider such an experiment. Future work will consider this possibility (also, see~\cite{Gondolo:2002np}).

We note that the anisotropy at radii beyond about the scale radius is sensitive to the shape of the PPSD profile and to the other parameters, and it can also be quite large. However, it seems unlikely that this grants a viable observational effect, since the density is so low there and substructure would dominate any emission.

The most novel result of this work is the constraint on the velocity dispersion profile. Requiring the anisotropy parameter to be physical (no greater than one) inside the virial radius implies a maximum value for the local total velocity dispersion of about $300\,\mathrm{km/s}$ or so. Typical assumed values for the local velocity dispersion (such as in the Standard Halo Model, $220\,\mathrm{km/s}$) do not seem to be in great danger, but this consistency check should be remembered in future model-building.

\acknowledgments
The author is grateful to Francesc Ferrer and Stanley D. Hunter for advice and discussions.
The author also greatly appreciates the anonymous referee's time and very helpful comments.
This work was supported by the U.S. DOE at Washington University in St. Louis.

\appendix
\section{Detailed expressions}\label{sec:details}
Here we discuss the derivation of the anisotropy parameter and related quantities in detail. We consider two-power-law density profiles, with inner slope $\gamma_0$ and outer slope $\gamma_\infty$,
\begin{equation}\label{eq:general_density}
\rho_{\gamma_0\gamma_\infty}\left( x\right) = \rho_s x^{-\gamma_0} \left( 1+x\right)^{-\gamma_\infty+\gamma_0}.
\end{equation}
The contained mass is, omitting the constant factor $4\pi \rho_s r_s^3$,
\begin{equation}
M_{\gamma_0\gamma_\infty}\left( x\right) = (-1)^{\gamma_0-1} B_{-x}\left( 3-\gamma_0, 1+\gamma_0-\gamma_\infty \right),
\end{equation}
where $B_z(a,b)$ is the incomplete beta function. Note the following particular cases:
\begin{align}
M_{1\gamma_\infty}(x) & = \frac{1 - \left[1 - x\left(2-\gamma_\infty\right)\right](1+x)^{2-\gamma_\infty} }{(3-\gamma_\infty)(2-\gamma_\infty)}, \\
M_{12}(x) & = x - \log(1+x),\\
M_\mathrm{NFW}(x) = M_{13}(x) & = -\frac{x}{1+x} + \log(1+x),\\
M_{\gamma_04}(x) & = \frac{1}{3-\gamma_0}\left(\frac{x}{1+x}\right)^{3-\gamma_0}.
\end{align}
From the PPSD power-law in \myeqref{ppsd} and the general density profile in \myeqref{general_density}, we have the radial velocity dispersion
\begin{equation}
\sigr^2(x) = \sigrs^2 \left[ x^{-\gamma_0+\ppsdnegslope} \left(\frac{2}{1+x}\right)^{\gamma_\infty-\gamma_0} \right]^{2/3}.
\end{equation}
For the NFW profile this is
\begin{equation}
\sigr^2(x) = \sigrs^2\left(\frac{4 x^{-1+\ppsdnegslope}}{(1+x)^2}\right)^{2/3}.
\label{eq:sigr}
\end{equation}
The expression for the anisotropy parameter $\ani\left(x\right)$ in \myeqref{ani} is general.
Specific to the case of the NFW profile, it is
\begin{equation}\label{eq:nfw_ani}
\ani(x) = \frac{5+15x}{6+6x} - \frac{\ppsdnegslope}{3} - \Sigma^{-2} \cdot 3 x^{-(2\ppsdnegslope +1)/3} \left(\frac{1+x}{2}\right)^{1/3} \left[-x + (1+x)\log( 1+x )\right].
\end{equation}
with $\Sigma^2 \equiv \sigrs^2/(4\pi G \rs^2 \rhos /3)$.
The upper limit on $\Sigma$ in the case of a NFW profile is
\begin{equation}
\Sigma^2 \le \Sigma^2_\mathrm{max} \equiv \frac{3^2 \cdot 2^{2/3}(1+c)^{4/3}\left[-c+(1+c)\log( 1+c)\right]}{c^{(1+2\ppsdnegslope)/3} \left[9c - 2\ppsdnegslope(1+c) - 1\right]}.
\label{eq:Sigma2_upper_limit}
\end{equation}
This translates to the upper limits on $\sigrs$ and $\sigrlocal$:
\begin{eqnarray}
\sigrs^2 \le \sigma^2_\mathrm{r,s,max} &\equiv& (4\pi G \rs^2\rhos /3) \Sigma^2_\mathrm{max}\left(\ppsdnegslope,c\right),\label{eq:sigrs_max}\\
\sigrlocal^2 \le \sigma^2_{\mathrm{r},\odot,\mathrm{max}} &\equiv& \left(\frac{4\,\xlocal^{-1+\ppsdnegslope}}{\left(1+\xlocal\right)^2}\right)^{2/3} \sigma^2_\mathrm{r,s,max}\left(\ppsdnegslope,\Mvir,\rs,c\right).\label{eq:sigrlocal_max}
\end{eqnarray}
Finally, because $\sigtlocal$ increases monotonically with $\sigrlocal$, its upper limit is
\begin{equation}\label{eq:sigtlocal_max}
\sigtlocal^2 \le (3 - 2\anilocal) \sigma^2_{\mathrm{r},\odot,\mathrm{max}},
\end{equation}
which depends on~$\ppsdnegslope$,~$\Mvir$,~$\rs$,~$c$, and~$\rlocal$.

\section{Asymptotic behavior}\label{sec:ani_limits}
We split the function for the NFW anisotropy parameter in \myeqref{nfw_ani} into two parts, so $\ani(x) = A(x) + B(x)$, with
\begin{eqnarray}
A(x) & = & \frac{5+15x}{6+6x} - \frac{\ppsdnegslope}{3}, \\
B(x) & = & -\Sigma^{-2}\cdot 3x^{-(2\ppsdnegslope +1)/3} \left(\frac{1+x}{2}\right)^{1/3} \left[-x + (1+x)\log( 1+x )\right].
\end{eqnarray}
The first part has simple asymptotic limits
\begin{equation}
A(x) \rightarrow \begin{cases}
  (5-2\ppsdnegslope)/6 & \text{for}\hspace{6pt}x\rightarrow 0\\
  (15-2\ppsdnegslope)/6 & \text{for}\hspace{6pt}x\rightarrow \infty
\end{cases}
\label{eq:Aasym}
\end{equation}
while the second is more complicated. In the limit $x \rightarrow 0$, we have
\begin{equation}
B(x) \rightarrow \begin{cases}
  -\infty & \text{if}\hspace{6pt}\ppsdnegslope > 5/2\\
  - 2^{-4/3}\cdot 3 \times \Sigma^{-2} & \text{if}\hspace{6pt}\ppsdnegslope=5/2\\
  0 & \text{if}\hspace{6pt}\ppsdnegslope<5/2
\end{cases}
\label{eq:Basym0}
\end{equation}
and in the limit $x \rightarrow \infty$, we have
\begin{equation}
B(x) \rightarrow \begin{cases}
  0 & \text{if}\hspace{6pt}\ppsdnegslope > 3/2\\
  -\infty & \text{if}\hspace{6pt}\ppsdnegslope\le 3/2
\end{cases}
\label{eq:BasymInf}
\end{equation}
As long as $3/2 < \ppsdnegslope < 5/2$, the extreme values of $\ani(x)$ are determined solely by $\ppsdnegslope$.

\section{Details of anisotropic velocity distributions}
\label{sec:vd_details}
\begin{figure}[t]
        \centering
        \includegraphics[width=\textwidth]{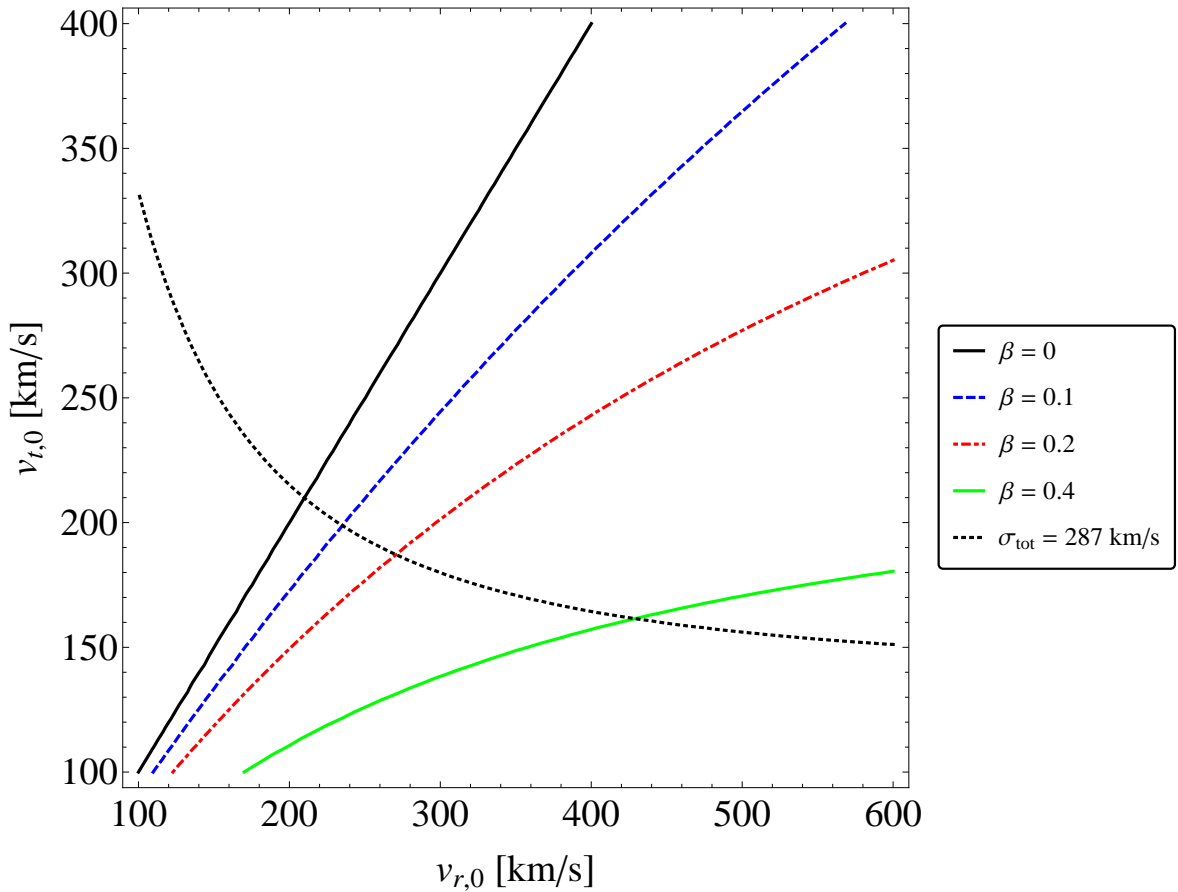}
        \caption{Contours of the parameters $\vrp$ and $\vtp$ that give the specified values of the anisotropy parameter or total velocity dispersion.}
        \label{fig:mao_contours}
\end{figure}
We use the velocity distribution in \myeqref{animao} to model the local velocity distribution, with $p=1.5$ from~\cite{Kuhlen:2013tra} and with $\vesc = 550.7\,\mathrm{km/s}$ from~\cite{Catena:2011kv}. The choice of parameters $\vrp$ and $\vtp$ determine the velocity dispersion and anisotropy parameter. Figure~\ref{fig:mao_contours} plots contours that give the specified value of $\ani$ or $\sigt$. In this work we choose $\sigt = 287\,\mathrm{km/s}$ as the mean value~\cite{Catena:2011kv}. For the isotropic case, this implies $\vrp = \vtp = 209.8\,\mathrm{km/s}$; for the anisotropic case, with $\ani = 0.2$, this implies $\vrp = 270.4\,\mathrm{km/s}$ and $\vtp = 187.1\,\mathrm{km/s}$.


\bibliographystyle{JHEP}
\bibliography{ppsd}

\providecommand{\href}[2]{#2}\begingroup\raggedright\begin{thebibliography}{10}

\bibitem{Bertone:2004pz}
G.~Bertone, D.~Hooper, and J.~Silk, {\it {Particle dark matter: Evidence,
  candidates and constraints}},  {\em Phys.Rept.} {\bf 405} (2005) 279--390,
  [\href{http://xxx.lanl.gov/abs/hep-ph/0404175}{{\tt hep-ph/0404175}}].

\bibitem{Robertson:2009bh}
B.~Robertson and A.~Zentner, {\it {Dark Matter Annihilation Rates with
  Velocity-Dependent Annihilation Cross Sections}},  {\em Phys.Rev.} {\bf D79}
  (2009) 083525, [\href{http://xxx.lanl.gov/abs/0902.0362}{{\tt
  arXiv:0902.0362}}].

\bibitem{Campbell:2010xc}
S.~Campbell, B.~Dutta, and E.~Komatsu, {\it {Effects of Velocity-Dependent Dark
  Matter Annihilation on the Energy Spectrum of the Extragalactic Gamma-ray
  Background}},  {\em Phys.Rev.} {\bf D82} (2010) 095007,
  [\href{http://xxx.lanl.gov/abs/1009.3530}{{\tt arXiv:1009.3530}}].

\bibitem{Ferrer:2013cla}
F.~Ferrer and D.~R. Hunter, {\it {The impact of the phase-space density on the
  indirect detection of dark matter}},  {\em JCAP} {\bf 1309} (2013) 005,
  [\href{http://xxx.lanl.gov/abs/1306.6586}{{\tt arXiv:1306.6586}}].

\bibitem{Gondolo:2002np}
P.~Gondolo, {\it {Recoil momentum spectrum in directional dark matter
  detectors}},  {\em Phys.Rev.} {\bf D66} (2002) 103513,
  [\href{http://xxx.lanl.gov/abs/hep-ph/0209110}{{\tt hep-ph/0209110}}].

\bibitem{Strigari:2009zb}
L.~E. Strigari and R.~Trotta, {\it {Reconstructing WIMP Properties in Direct
  Detection Experiments Including Galactic Dark Matter Distribution
  Uncertainties}},  {\em JCAP} {\bf 0911} (2009) 019,
  [\href{http://xxx.lanl.gov/abs/0906.5361}{{\tt arXiv:0906.5361}}].

\bibitem{Catena:2011kv}
R.~Catena and P.~Ullio, {\it {The local dark matter phase-space density and
  impact on WIMP direct detection}},  {\em JCAP} {\bf 1205} (2012) 005,
  [\href{http://xxx.lanl.gov/abs/1111.3556}{{\tt arXiv:1111.3556}}].

\bibitem{Frandsen:2011gi}
M.~T. Frandsen, F.~Kahlhoefer, C.~McCabe, S.~Sarkar, and K.~Schmidt-Hoberg,
  {\it {Resolving astrophysical uncertainties in dark matter direct
  detection}},  {\em JCAP} {\bf 1201} (2012) 024,
  [\href{http://xxx.lanl.gov/abs/1111.0292}{{\tt arXiv:1111.0292}}].

\bibitem{binney}
J.~{Binney} and S.~{Tremaine}, {\em {Galactic Dynamics: Second Edition}}.
\newblock Princeton University Press, 2008.

\bibitem{Osipkov:1979}
L.~P. {Osipkov}, {\it {Spherical systems of gravitating bodies with an
  ellipsoidal velocity distribution}},  {\em Soviet Astronomy Letters} {\bf 5}
  (1979) 42--44.

\bibitem{Cuddeford:1991}
P.~{Cuddeford}, {\it {An analytic inversion for anisotropic spherical
  galaxies}},  {\em \mnras} {\bf 253} (Dec., 1991) 414--426.

\bibitem{Gerhard:1991}
O.~E. {Gerhard}, {\it {A new family of distribution functions for spherical
  galaxies}},  {\em \mnras} {\bf 250} (June, 1991) 812--830.

\bibitem{Baes:2002tw}
M.~Baes and H.~Dejonghe, {\it {The Hernquist model revisited: Completely
  analytical anisotropic dynamical models}},  {\em Astron.Astrophys.} {\bf 393}
  (2002) 485--498, [\href{http://xxx.lanl.gov/abs/astro-ph/0207233}{{\tt
  astro-ph/0207233}}].

\bibitem{VanHese:2010qy}
E.~Van~Hese, M.~Baes, and H.~Dejonghe, {\it {On the universality of the global
  slope -- anisotropy inequality}},  {\em Astrophys.J.} {\bf 726} (2011) 80,
  [\href{http://xxx.lanl.gov/abs/1010.4301}{{\tt arXiv:1010.4301}}].

\bibitem{Zait:2007es}
A.~Zait, Y.~Hoffman, and I.~Shlosman, {\it {Dark Matter Halos: Velocity
  Anisotropy -- Density Slope Relation}},
  \href{http://xxx.lanl.gov/abs/0711.3791}{{\tt arXiv:0711.3791}}.

\bibitem{Sparre:2012zk}
M.~Sparre and S.~H. Hansen, {\it {The behaviour of shape and velocity
  anisotropy in dark matter haloes}},  {\em JCAP} {\bf 1210} (2012) 049,
  [\href{http://xxx.lanl.gov/abs/1210.2392}{{\tt arXiv:1210.2392}}].

\bibitem{Wojtak:2013eia}
R.~Wojtak, S.~Gottloeber, and A.~Klypin, {\it {Orbital anisotropy in
  cosmological haloes revisited}},
  \href{http://xxx.lanl.gov/abs/1303.2056}{{\tt arXiv:1303.2056}}.

\bibitem{Taylor:2001bq}
J.~E. Taylor and J.~F. Navarro, {\it {The Phase - space density profiles of
  cold dark matter halos}},  {\em Astrophys.J.} {\bf 563} (2001) 483--488,
  [\href{http://xxx.lanl.gov/abs/astro-ph/0104002}{{\tt astro-ph/0104002}}].

\bibitem{VanHese:2008ce}
E.~Van~Hese, M.~Baes, and H.~Dejonghe, {\it {The dynamical structure of dark
  matter halos with universal properties}},  {\em Astrophys.J.} {\bf 690}
  (2009) 1280--1291, [\href{http://xxx.lanl.gov/abs/0809.0901}{{\tt
  arXiv:0809.0901}}].

\bibitem{Ma:2009ek}
C.-P. Ma, P.~Chang, and J.~Zhang, {\it {Is the Radial Profile of the
  Phase-Space Density of Dark Matter Halos a Power-Law?}},
  \href{http://xxx.lanl.gov/abs/0907.3144}{{\tt arXiv:0907.3144}}.

\bibitem{Hansen:2004gs}
S.~H. Hansen, {\it {Dark matter density profiles from the jeans equation}},
  {\em Mon.Not.Roy.Astron.Soc.} {\bf 352} (2004) L41,
  [\href{http://xxx.lanl.gov/abs/astro-ph/0405371}{{\tt astro-ph/0405371}}].

\bibitem{Dehnen:2005cu}
W.~Dehnen and D.~McLaughlin, {\it {Dynamical insight into dark-matter haloes}},
   {\em Mon.Not.Roy.Astron.Soc.} {\bf 363} (2005) 1057--1068,
  [\href{http://xxx.lanl.gov/abs/astro-ph/0506528}{{\tt astro-ph/0506528}}].

\bibitem{Austin:2005ks}
C.~G. Austin, L.~L. Williams, E.~I. Barnes, A.~Babul, and J.~J. Dalcanton, {\it
  {Semi-analytical dark matter halos and the Jeans equation}},  {\em
  Astrophys.J.} {\bf 634} (2005) 756--774,
  [\href{http://xxx.lanl.gov/abs/astro-ph/0506571}{{\tt astro-ph/0506571}}].

\bibitem{Navarro:1996gj}
J.~F. Navarro, C.~S. Frenk, and S.~D. White, {\it {A Universal density profile
  from hierarchical clustering}},  {\em Astrophys.J.} {\bf 490} (1997)
  493--508, [\href{http://xxx.lanl.gov/abs/astro-ph/9611107}{{\tt
  astro-ph/9611107}}].

\bibitem{Schmidt:2009kz}
K.~B. Schmidt, S.~H. Hansen, J.~H. An, L.~L. Williams, and A.~V. Macci'o, {\it
  {Dark Matter Angular Momentum Profile from the Jeans Equation}},  {\em
  Astrophys.J.} {\bf 694} (2009) 893--901,
  [\href{http://xxx.lanl.gov/abs/0901.0928}{{\tt arXiv:0901.0928}}].

\bibitem{Catena:2009mf}
R.~Catena and P.~Ullio, {\it {A novel determination of the local dark matter
  density}},  {\em JCAP} {\bf 1008} (2010) 004,
  [\href{http://xxx.lanl.gov/abs/0907.0018}{{\tt arXiv:0907.0018}}].

\bibitem{Hansen:2004qs}
S.~H. Hansen and B.~Moore, {\it {A Universal density slope - velocity
  anisotropy relation for relaxed structures}},  {\em New Astron.} {\bf 11}
  (2006) 333, [\href{http://xxx.lanl.gov/abs/astro-ph/0411473}{{\tt
  astro-ph/0411473}}].

\bibitem{An:2009nc}
J.~H. An and N.~W. Evans, {\it {A theorem on central velocity dispersions}},
  {\em Astrophys.J.} {\bf 701} (2009) 1500--1505,
  [\href{http://xxx.lanl.gov/abs/0906.3673}{{\tt arXiv:0906.3673}}].

\bibitem{Ludlow:2011cs}
A.~D. Ludlow, J.~F. Navarro, M.~Boylan-Kolchin, V.~Springel, A.~Jenkins,
  et~al., {\it {The density and pseudo-phase-space density profiles of cold
  dark matter haloes}},  {\em Mon.Not.Roy.Astron.Soc.} {\bf 415} (2011)
  3895--3902, [\href{http://xxx.lanl.gov/abs/1102.0002}{{\tt
  arXiv:1102.0002}}].

\bibitem{Ullio:2000bf}
P.~Ullio and M.~Kamionkowski, {\it {Velocity distributions and annual
  modulation signatures of weakly interacting massive particles}},  {\em JHEP}
  {\bf 0103} (2001) 049, [\href{http://xxx.lanl.gov/abs/hep-ph/0006183}{{\tt
  hep-ph/0006183}}].

\bibitem{Vogelsberger:2008qb}
M.~Vogelsberger, A.~Helmi, V.~Springel, S.~D. White, J.~Wang, et~al., {\it
  {Phase-space structure in the local dark matter distribution and its
  signature in direct detection experiments}},  {\em Mon.Not.Roy.Astron.Soc.}
  {\bf 395} (2009) 797--811, [\href{http://xxx.lanl.gov/abs/0812.0362}{{\tt
  arXiv:0812.0362}}].

\bibitem{Mao:2012hf}
Y.-Y. Mao, L.~E. Strigari, R.~H. Wechsler, H.-Y. Wu, and O.~Hahn, {\it
  {Halo-to-Halo Similarity and Scatter in the Velocity Distribution of Dark
  Matter}},  {\em Astrophys.J.} {\bf 764} (2013) 35,
  [\href{http://xxx.lanl.gov/abs/1210.2721}{{\tt arXiv:1210.2721}}].

\bibitem{Kuhlen:2013tra}
M.~Kuhlen, A.~Pillepich, J.~Guedes, and P.~Madau, {\it {The Distribution of
  Dark Matter in the Milky Way's Disk}},
  \href{http://xxx.lanl.gov/abs/1308.1703}{{\tt arXiv:1308.1703}}.

\bibitem{Bozorgnia:2013pua}
N.~Bozorgnia, R.~Catena, and T.~Schwetz, {\it {Anisotropic dark matter
  distribution functions and impact on WIMP direct detection}},
  \href{http://xxx.lanl.gov/abs/1310.0468}{{\tt arXiv:1310.0468}}.

\bibitem{Gondolo:2012rs}
P.~Gondolo and G.~B. Gelmini, {\it {Halo independent comparison of direct dark
  matter detection data}},  {\em JCAP} {\bf 1212} (2012) 015,
  [\href{http://xxx.lanl.gov/abs/1202.6359}{{\tt arXiv:1202.6359}}].

\end{thebibliography}\endgroup

\end{document}